\documentclass[11pt,reqno]{amsart}
\usepackage[colorlinks=true, linkcolor=blue, urlcolor=blue, citecolor=blue]{hyperref}
\usepackage{natbib}
\usepackage{graphicx}
\usepackage{xcolor}
\usepackage{geometry}                
\usepackage{longtable,booktabs}
\usepackage{rotating}
\usepackage{subfig}

\usepackage{amsmath,amssymb,amsthm}

\theoremstyle{plain}
\long\def\comment#1{}
\newtheorem{algorithm}{Algorithm}

\theoremstyle{definition}

\numberwithin{definition}{section}

\newtheorem{example}{Example}
\newtheorem{remark}{Comment}

\newcommand{\g}{g}

\renewcommand{\sc}{\mathsf{c}}

\newcommand{\argmin}{\mathop{\mathrm{arg\,min}{}}}

\let\hat\widehat

\definecolor{DSgray}{cmyk}{0,1,0,0}

\begin{document}
\begin{center}

\end{center}

\title[Adaptive Discrete Smoothing]{Adaptive Discrete Smoothing for High-Dimensional and Nonlinear Panel Data}
\author{Xi Chen, Ye Luo, and Martin Spindler}\thanks{Luo: Hong Kong University, Faculty of Business and Economics, kurtluo@gmail.com, Chen: New York University, Stern School of Business, xchen3@stern.nyu.edu,
Spindler: University of Hamburg and Max Planck Society, martin.spindler@uni-hamburg.de. We thank the German Research Foundation (DFG) for financial support through a grant for \textquotedblleft Initiation of International Collaborations\textquotedblright. We thank Chunrong Ai, Christoph Breunig, Chris Hansen, Jerry Hausman, Whitney Newey, Zhentao Shi, Frank Windmeijer and seminar participants at the University of Florida and University of Munich for helpful comments.}
\date{December, 2019.}
\maketitle

\begin{footnotesize}
\textbf{Abstract.}


In this paper we develop a data-driven smoothing technique for high-dimensional and non-linear panel data models. We allow for individual specific (non-linear) functions and estimation with econometric or machine learning methods by using weighted observations from other individuals. The weights are determined by a data-driven way and depend on the similarity between the corresponding functions and are measured based on initial estimates. The key feature of such a procedure is that it clusters individuals based on the distance / similarity between them, estimated in a first stage. Our estimation method can be combined with various statistical estimation procedures, in particular modern machine learning methods which are in particular fruitful in the high-dimensional case and with complex, heterogeneous data. The approach can be interpreted as a \textquotedblleft soft-clustering\textquotedblright\ in comparison to traditional\textquotedblleft\ hard clustering\textquotedblright that assigns each individual to exactly one group. We conduct a simulation study which shows that the prediction can be greatly improved by using our estimator. Finally, we analyze a big data set from didichuxing.com, a leading company in transportation industry, to analyze and predict the gap between supply and demand based on a large set of covariates. Our estimator clearly performs much better in out-of-sample prediction compared to existing linear panel data estimators.



\end{footnotesize}

\section{Introduction}
\subsection{Motivation}
Panel or longitudinal data are very important tool for empirical research in economics, social sciences, biostatistics and many other fields. Most of the literature on non- and semi-parametric panel models is based on the assumption
that the regression function is the same across individuals. But this assumption, however, could be unrealistic in many applications, in particular when the number of observed individuals is large or there is unobserved heterogeneity across the individuals. In modern internet data with hundreds of thousands of users and a relatively short longitudinal time frame, it is important to study complicated and non-linear heterogeneity among different individuals.


A classical example is the prediction of demand and supply. More specifically, we will consider data from didichuxing.com (the Chinese counterpart of Uber in the U.S.) and estimate gaps between real time supply and demand in a metropolitan area in China being partitioned into 66 districts and the 24 hours day time being divided into 144 intervals, each containing 10 minutes. For companies like didichuxing.com it is crucial to develop a good model for estimating supply and demand gaps to provide better dispatching and services and to develop a dynamic pricing strategy.

Obviously, the model for predicting the demand and supply of taxi service at a given location and a specified time spot should not be uniform over all time-location combinations. Even at the same time spot, e.g., 8:00pm - 8:10pm, the supply, the demand, and the gap between them, as well as traffic conditions, vary dramatically across different locations in the city. Also, for the same location, the supply and demand are quite different in different time spots.

For modeling this kind of data sets, panel data models are widely used both in the literature but also in many empirical applications. However, due to significant heterogeneity across time and locations, the linear panel data model works poorly in terms of predicting the supply and demands, both in and out of sample when the number of observations per unit is relatively small. From a theoretical point of view, nonparametric panel data models are still a challenge as usually there are only a limited number of observations per individual available which makes nonparametric estimation of the individual regression functions quite imprecise or even impossible.
Despite these challenges, one can often observe a certain \emph{grouping structure} among models for different individuals in real data sets. For example, it is possible that district number one, at 8am, may share the same or a similar model for the gap between supply and demand, with district two at 6pm. One one hand, such a grouping structure among individual models can be used to improve the prediction for each model by borrowing observations from other similar models. On the other hand, these latent group structure among models is unknown and can hardly be figured out using common knowledge.


In this paper we propose a new adaptive discrete-smoothing (ADS) method which is designed to utilize the latent grouping structure among a large number of nonparametric panel data models. More precisely, the proposed ADS method simultaneously ``clusters'' the cross-sectional individuals and conducts model estimation.

In a non-parametric panel data model with the fixed effect setting, we assume that there are individual fixed effects $\alpha_i$ which determine the nonparametric regression function of each individual.  As we explain later in the model specification part, these individual effects can be interpreted as the indices of a class of functions that determine the shape of the individual regression functions.
To characterize the grouping structure among regression functions,  
we make an essential continuity/smoothness assumption: individuals with close values of $\alpha$s have similar regression functions. This assumption enables us to effectively address the curse of dimensionality (i.e., the insufficiency of samples for each individual model) in estimating nonparametric panel data models. The main idea behind the construction of our estimator is to use all the observations for estimating each individual model, including the observations from other individuals, but these observations receive a lower weight in estimation. More specifically, we develop a general two-stage estimator, where the first stage estimates the weights that characterize the ``similarity'' of the underlying regression functions. Given the estimated weight matrix between pairs of individuals, the second stage estimates the regression function for each individual using all the weighted observations.

Our idea shares a similar spirit as the ``discrete smoothing'' idea from \cite{LiRacine}.
In particular, \cite{LiRacine} studied the nonparametric regression with discrete and continuous variables and used fixed weights for observations falling in other \textquotedblleft cells\textquotedblright, where each cell corresponds to a combination of categorical variables (\textquotedblleft discrete smoothing\textquotedblright). They showed that the mean squared error can be effectively reduced by using the weighted observations across cells. Instead of using the fixed weights, our adaptive smoothing approach over discrete cells depends on how similar the corresponding underlying regressions functions are. To accomplish this, we introduce data-driven weights that measure the similarity (distance) of regression functions across different individuals. Using observations from other individuals, on the one hand, introduces bias, but on the other hand, adding additional observations in the estimation process might reduce variance. By constructing the data-driven weights, our method leads to a better variance-bias trade-off and hence faster convergence rates in theory.


Generally speaking, the proposed methodology is applicable to almost any statistical procedure which allows weighted estimation, including kernel estimation, series / sieves, and modern machine learning methods like Lasso, Boosting and many others methods. As the weights employed for the smoothing of the regression functions are data-driven, we call our method \textquotedblleft \emph{adaptive discrete smoothing}\textquotedblleft, or \emph{ADS} in the rest of the paper. Our  ADS method applies to both discrete and continuous indices $\alpha$ without having any prior knowledge of $\alpha$. As the rise of digitization in many fields leads to complex, high-dimensional data sets, we focus in the this paper on Lasso.
In this paper, we focus on panel data models. The methods can also be used for non-parametric regression with both continuous and discrete variables, as considered in \cite{LiRacine}. Adaptive discrete smoothing for non-parametric regression is considered in an accompanying paper (\cite{CLS2020}). For the theoretical properties of our estimator first results are available for some estimation methods upon request.\footnote{An updated version with theoretical results is work in progress.} 


\textbf{Outline}
In Section 2 we give a heuristic motivation for our procedure. In Section 3 the ADS procedure is formally introduced in a general setting and specified for relevant problems. Section 4 presents a set of simulation studies to demonstrate the power of our methods compared to other alternatives. In Section 5, we present an empirical application: a study on didichuxing.com's panel data for predicting gap between supply and demand. Finally, we conclude.

\subsection{Related Literature}
Since the use of panel data is popular in many disciplines such as social, medical, financial studies, the panel data model has always been an active research area (see, e.g., \cite{Baltagi1992,Wooldbridge:10,Hsiao:14}). Due to the large volume of work in this area, we only provide a brief survey of closely related works.

Our paper is related to different lines of recent research on panel data model. The first line is high-dimensional (or diverging dimensional) panel data model (see, e.g., \cite{Kock:13,kock:16:panel,BCHK:FE,zhu2017}).  In addition to parameter estimation and inference,  \cite{Li:16:break} and \cite{QianSu:16} further study the estimation of common structural breaks in linear panel data model in the diverging dimensional case. 

Moreover, our work extends also non-parametric panel methods and recent approaches employing clustering for estimation of panel data models. Nonlinear panel data have been a field of active research. A comprehensive survey is given in \cite{AB:2011}. Most of the literature on non- and semi-parametric panel models is based on the assumption that the regression function is the same across individuals; see \cite{Henderson}, \cite{Mammen} and \cite{Qian:03} among many others.
As argued above, this assumption might not be realistic realistic in many applications.
To relax the assumption, recent research has focused on assuming a group structure and employing cluster methods for panel data models. This means that every individual belongs to a group and the group assignment is unknown and has to be estimated. In such a setting, the number of groups usually needs to be finite and predefined. \cite{VL:2017} considered a nonparametric regression model that the observed individuals can be grouped into a number of classes whose members all share the same regression function. This is a special case of the model in Example \ref{example:non-linear}  with discrete indices (see Section \ref{example:non-linear}).  They develop a statistical procedure to estimate the unknown group structure  and then estimate each regression function by averaging the individual functional estimates within each group. \cite{Bonhomme:2015}  introduce time-varying grouped patterns of heterogeneity in linear panel data models. 
The parameters are estimated using a grouped fixed effects estimator that minimizes a least squares criterion with respect to all possible groupings of the cross-sectional units based on $K$-means clustering.  \cite{SLM2016}  develop two-step and iterative panel data estimators based on a discretization of unobserved heterogeneity employing clustering. \cite{SSP:2016} provide a novel mechanism for identifying and estimating latent group structures in panel data using penalized techniques. They consider both linear and nonlinear models where the regression coefficients are heterogeneous across groups but homogeneous within a group and the group membership is unknown. \cite{SSP:2016} proposes a novel penalty term called classifier-Lasso ($C$-Lasso). More specially, assuming that the true regression coefficients for $N$ individuals $\{\beta_1^*, \ldots, \beta_N^*\}$ only take $K_0$ distinct vector values, the  $C$-Lasso imposes the non-convex penalty $\sum_{i=1}^N \prod_{k=1}^{K_0} \|\beta_i-\alpha_k\|_2$, where both $\{\beta_i\}_{i=1}^N$ and $\{\alpha_k\}_{k=1}^{K_0}$ are the decision variables in the optimization. This mixed additive-multiplication form of  penalty tries to shrink individual regression coefficients $\beta_i$ to an unknown group-level coefficient vector $\alpha_k$.

Despite the popularity of investigating the grouping effect, there are several limitations in the existing approaches. First, the computation of the estimators is usually quite demanding. For example, some estimators rely on the $k$-means algorithm to learn the group structure. However, finding the optimal solution to the $k$-means clustering problem is known to be NP-hard even for two clusters \citep{Aloise:09}. Other estimators either depend on the exhaustive search of the group structure or involve solving some non-convex optimization problems for identifying the group structure. Second, the estimators with an explicit clustering step usually require the number of clusters is known a priori.  Approaches have been developed to determine the unknown number of groups but this is again computational demanding and might need additional assumptions. Third, it is very reasonable that in many applications there are not only group effects, but also individual effects. There might not be a hard group structure among different individuals. For example, some individuals might have similar coefficients but do not share exactly the same model. Finally, except for \cite{VL:2017}, most existing research mainly focuses on linear or parametric panel data models and does not deal with general nonparametric regression functions.

To address these challenges, we propose a unified approach for a wide range of panel data models that include both parametric and non-parametric models, continuous and discrete variables, fixed-dimensional and high-dimensional settings. Instead of enforcing the ``hard clustering'' that assigns each individual to exactly one group, our  approach can be interpreted as a \textquotedblleft soft clustering\textquotedblright with weights determined by the similarities between groups. This soft clustering approach is not only computationally attractive as compared to those ``hard clustering'' methods but also avoids fixing the number of groups prior to that.

\subsection{Important Examples in Econometrics}
\label{sec:example}

\begin{example}[Non-linear Panel data]\label{example:non-linear}
We assume that
\begin{equation}
y_{it}=f(x_{it}, \alpha_i) + \varepsilon_{it},
\qquad \mathbb{E}[\varepsilon_{it}|x_{it}, \alpha_i]=0.
\end{equation}
$\varepsilon_{it}$ denotes the error term which is contemporaneously uncorrelated with the regressors $x_{it}$.

The dependent variable depends nonparametrically on the regressors and an unobserved \textquotedblleft fixed effect\textquotedblright $\alpha_i$. So each individual $i$ might be subject to a different, nonlinear function $f_{i} \equiv f(\ \cdot\ ,\alpha_i)$ which is unknown and has to be estimated. Different values of $\alpha_i$ lead to different functions $f_{i}$. The effects $\alpha_i$ are not identified, but we show that the functions $\{f_i\}$ represent the unobserved heterogeneity $\alpha_i$. One way to interpret this is that the effects $\alpha_i$ simply serve as an index for the nonparametric functions and different $\alpha_i$ lead to different regression functions $f_i$. Without loss of generality we assume that $\alpha \in [0,1]$ and define the family of functions that are indexed by $\alpha$, namely $\mathcal{F}$. Formally, $\mathcal{F}:=\{g|g_{\alpha}(x):=f(x,\alpha),\alpha\in [0,1]\}$. For each individual $i=1,2,...,N$, the fixed effect $\alpha_i$ leads to a response function $f_i(x):=f(x,\alpha_i)\in \mathcal{F}$. When $\{\alpha_1, \ldots, \alpha_n\}$ take a finite discrete values, this general model includes the models considered in \cite{VL:2017} as special cases. The core idea of our estimator relies on a continuity assumptions: individuals with similar values of $\alpha$ also have similar regression functions. This assumption will be stated more rigorously in the next sections. In the following comment we will also argue that non-parametric functions allow a linear high-dimensional approximation which is often sparse. As our interest is mainly on modern high-dimensional data sets we will mainly focus on the high-dimensional setting with keeping in mind that non-linear models might be represented as a high-dimensional problem.
\end{example}

\begin{remark}{Approximate Sparse Models}\label{ASM}
We start with a nonlinear relationship of the form
\[ y_i=f(z_i) + \varepsilon_i, \varepsilon_i \sim N(0, \sigma^2), i=1,\ldots,n,\]
where $y_i$ is the outcome variable, $z_i$ is a $k_z$-vector of elementary regressors, $f(z_i)$ is the regression function, and $\varepsilon_i$ are $i.i.d$ disturbances. Let $x_i=P(z_i)$, where $P(z_i)$ is a vector of dimension $p=p_n$, that contains a dictionary of possibly technical transformations of $z_i$, including a constant. The values $x_1,\ldots, x_n$ are treated fixed, and normalized.  The regression function $f(z_i)$ admits an approximate sparse form, if there exists $\beta_0$ such that
\[
f(z_i)=x_i' \beta_0 + r_i, ||\beta_0||_0 \leq s, c_s:=\{\mathbf{E}_n[r_i^2]\}^{1/2} \leq K \sigma \sqrt{s/n},
\]
where $s=s_n=o(n/ \log p)$ and $K$ is a constant independent of $n$.
\end{remark}

The methodology we introduce can be applied directly for nonparametric panel models, e.g. employing kernel or Sieves methods. But in the rest of the paper we will focus on linear and high-dimensional panel data models. First, modern data sets are often high-dimensional and of particular interest. Second, non-parametric models can often be represented as approximate sparse models as defined in the remark above.

\begin{example}[Linear panel data with heterogeneous coefficients]\label{example:linear}
We assume that
\begin{equation}\label{eq:Lasso}
y_{it}=x_{it}\beta(\alpha_i)+ \varepsilon_{it}, \qquad \mathbb{E}[\varepsilon_{it}|x_{it}, \alpha_i]=0,
\end{equation}
where $\beta(\cdot)\in \mathbb{R}^p$ is a link function that takes $\alpha_i$ as an input. That said, $\beta(\alpha_i)$ is a $p$-dimensional vector indexed by $\alpha_i\in [0,1]^{d_\alpha}$, where in the high dimensional case, $d_\alpha$ is much smaller than $\min(N, p)$, and $f_i(x_{it})=x_{it}\beta(\alpha_i)$.


When $p$ is fixed, we can simply estimate the model by linear regressions. It implies that we will be running $N$ linear regressions, one for each individual. When $p$ is high-dimensional, i.e., $p>>T$, we can estimate such a model by different machine learning techniques, such as Lasso, and $L_2$-Boosting.





\end{example}

\section{Motivation and Heuristic Derivation}

\begin{figure}%
\includegraphics[width=0.33\textwidth]{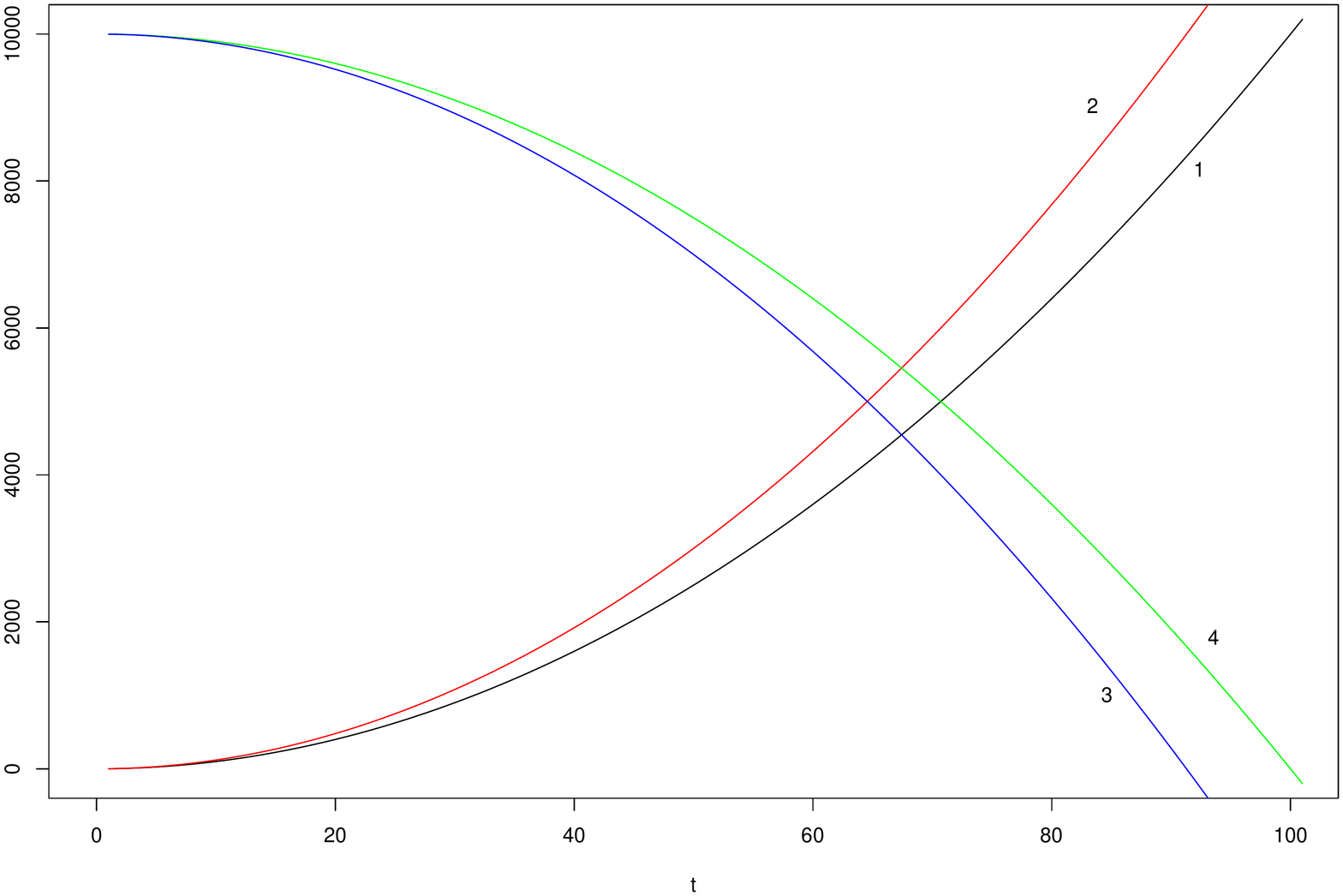}%
\caption{}%
\label{fig:panel}%
\end{figure}

To illustrate the core of our idea, we consider estimation of (nonparametric) panel functions / longitudinal data. Figure \ref{fig:panel} shows for illustration individual specific functions for four individual.  A naive approach might be to estimate a non-parametric function for each individual separately. But if the number of observations per individual $T$ is small, the regression functions cannot be estimated precisely. It seems that the regression functions for individuals 1 and 2 and for individuals 3 and 4 are similar. Hence, for estimation of the regression function of individual $1$ it is reasonable to use (\textquotedblleft borrow\textquotedblright) observations from individual $2$ for estimation, maybe with some lower weight. This might introduce bias, but decrease variance, leading to an increased MSE. Our idea is now to use data-dependent weights $\lambda$ and the weights reflect the similarity between the curves of all individuals. If the curves for individuals, here for example 1 and 2, are similar the weights should be close to one, as the information in the observations for individual 2 is valuable for estimation the function of individual 1. If the curves are very different, like the functions for 1 and 3, the weights should be small or zero. In the adaptive discrete smoothing procedure we propose the weights are based on the similarity between the two curves. The similarity is measured by a distance measure / metric  $\rho$ based on initial estimates, 
$\rho(\hat f_i(\cdot), \hat f_j(\cdot))$ where $\hat f_k, k=1,\ldots,N$ denote the estimated individual regression functions. The weight $\lambda$ is then given by the expression $\delta \exp(- \gamma \rho(\hat  f_i(\cdot), \hat  f_j(\cdot))$ with $\delta$ and $\gamma$ denoting constants. The weights for observation belonging to the individual itself are set to $1$. The weights are hence determined in data-driven / data-dependent way. Based on initial estimates for each individual, a weighting matrix is determined, measuring the similarity of functions. Finally, the individual functions are estimated using the weights. This procedure can be applied to all estimation methods which allow for weighted estimation. As for the initial estimates few observations might be available, but many potential covariates, we focus in this paper on Lasso estimation.

To make the idea more formal, we first introduce further notation. 
We consider panel data, in particular we observe $N$ individuals and for each individual we observe $T$ periods. The dependent variable is denoted as $y_i=(y_{i1}, \ldots, y_{iT})'$, $i=1,\ldots,n$ and the predictors are denoted as $x_i=(x_{i1},\ldots,x_{iT})^T$. The covariates $x_{it}$ are $p-$dimensional vectors $(x_{it,1}, \ldots, x_{it,p})'$.

Given any $i$, we assume that $f_i(x_{it}):=f(x_{it},\alpha_i)$ and $y_{it}=f(x_{it},\alpha_i) + \varepsilon_{it}$ for error terms $\varepsilon_{it}$. The $\alpha_i$ stands for the individual specific effect in the non-parametric panel data model. In general, the $\alpha_i$, serving as an low-dimensional index of $f_i$, but can be considered as multiple dimensional as well. In the panel data setting, $\alpha_i$ is usually consider as a scalar.

One naive way of predicting $y_{it}$ is through building individual-level models:

 $\hat f_i:=\argmin_{f\in \mathcal{F}} \sum_{t=1}^T (y_{it}-f(x_{it}))^2$, where $\mathcal{F}$ is a pre-specified functional space.

However, the naive estimator suffers from several drawbacks. First of all, in most of the real applications of panel data, we face the large $N$ and small $T$ problem, and estimating functions independently from other individual's observations lead to limited amount of data - only a sample size of $T$ for estimating each $f_i$.

In the traditional linear panel data model, $f_i(x_{it})=\alpha_i+x_{it}\beta+\varepsilon_{it}$, with $\mathbb{E}[\varepsilon_{it}|x_{it},\alpha_i]=0$. In such a model, the estimation process utilizes $N\times T$ samples at the same time because by assumption, $\beta$ is a common parameter across all individuals $i$. On contrast, the random coefficient model $f_i(x_{it}) = \alpha_i + x_{it}\beta_i+\varepsilon_{it}$ is a model where all $\beta_i$ are randomly distributed and can only be estimated by using individual longitudinal data without additional assumptions.

Suppose there exists group structures amongst all individuals in the panel data. One extreme case is that the number of such groups is only finite, which means that there are hidden labels that we do not observe, which we call them $\alpha_i$, $i=1,2,...,N$. When $\alpha_i=\alpha_j$, we say that the two individuals belong to the same group, and $f_i(x)=f(x,\alpha_i) =f(x,\alpha_j)=f_j(x)$. The linear panel data model is an special case because the coefficients $\beta$ are common across all individual $i$. A more general case is that we assume the distribution of $\alpha_i$ is continuous, and we allow that all individuals are different but might share some similarities. As $N$ becomes larger and larger, for any individual $i$, there must exist other individuals whose individual effects $\alpha$ are close enough to $\alpha_i$. Our idea is to assign proper weights to each individual $j\neq i$, and such weights should reflect the similarity between $\alpha_i$ and $\alpha_j$.

\section{General Discrete Smoothing Estimator for High-Dimensional and Non-Linear Panel Data Models}
\subsection{Generic ADS Algorithm for Panel Data}
In this section we introduce a generic algorithm for adaptive discrete smoothing of panel data. We consider the model  
$y_{it}=f(x_{it},\alpha_i) + \varepsilon_{it}, i = 1,\ldots,N$ and $t=1,\ldots T$. $x_{it}$ are $p-$dimensional vectors of regressors. As mentioned in the previous section the algorithm consists of three steps: First, providing initial estimates $\hat f_i(x_{it}), i=1,\ldots,N$. Second, constructing of the weighting matrix $W$. Third, for all $i=1,\ldots,N$: weighted estimation of $\tilde{f}_i(x_{it})$. The algorithm is very flexible and can be combined with any estimation methods that allows weighted estimation, including ordinary least squares, kernel regression, series regression, maximum likelihood estimation and modern machine learning methods like Lasso, Boosting, Neural Networks. The generic algorithm can be described as follows:

\begin{algorithm}[Generic ADS-Algorithm]\label{Algo:generic}
\begin{itemize}
    \item[(1)](First stage) Estimation of the first stage. Construction of initial estimates $\hat f^0_i, i=1,...\ldots,N$ using only the corresponding $T$ of each individual to construct the individual-specific first step estimators $\hat f^0_i$ 
    \item[(2)](Construction Weighting matrix) Compute $W$ matrix such that
    \begin{equation}\label{eq:Wkernel}
    W(i,j):=\delta \exp(-\gamma\rho(\hat f^0_i(x),\hat f_j^0(x))),
    \end{equation}
    for all $i\neq j$, where $\rho(\cdot,\cdot)$ is a metric. Set $W(i,i)=1$ for all $i=1,2,...,N$. 
    \item[(3)](Second stage) Weighted estimation of $f_i$ by using all observations with weights given by $W(i,j)$. The final estimator is denoted by $\widetilde{f}_i(x), i=1,\ldots,N.$
\end{itemize}
\end{algorithm}

\begin{remark}
 \begin{enumerate}
	 \item The choice of the metric / distance measure $\rho$ in Step 2 depends on the estimator. E.g. if ols or Lasso regression is employed, a natural choice is the Euclidean norm of the difference of the estimated coefficient vectors, $\rho(\hat \beta_i, \hat \beta_j)= ||\hat \beta_i - \hat \beta_j||^2_2$. In the case of kernel or series estimation, one might choose: $\rho(\hat f_i, \hat f_j)= \int |\hat f_i(x)-\hat f_j(x)|^2d\hat \mu(x)$ and $\hat \mu(x)$ as the empirical measure.
		\item In the last step (Second stage) a weighted regression is conducted. So for estimation all units are used, but with the weights employed in step 2. Hence, any estimation method supporting weights can be used with our method.
 \end{enumerate}
\end{remark}

\begin{remark}
A modification of the proposed algorithm is to iterate the calculation of the weights: After the third step the weights are updated with the estimated functions $\widetilde{f}_i(x), i=1,\ldots,N$ and then Step 3 is repeated. Either one stops then or the updating of the weights is repeated until the change of weights falls below some threshold.
\end{remark}

\begin{remark}[Parallel computation]
From the computation perspective, for both the first and second stage estimation, all $N$ estimators can be constructed in a fully parallel way, and thus the method is computationally attractive.
\end{remark}

\begin{remark}
In principle, the individual fixed effect $\alpha_i$ can be viewed as unknown categories. The ADS algorithm we propose has some similarity with clustering algorithms, e.g., $k-$ means algorithm, but offers more flexibility and has some advantages. First, the number of groups has not to be known or specified before. Rather, the grouping is determined in a data-driven fashion. The proposed estimator can cope with both discrete or \textquotedblleft continuous\textquotedblright categories. In Section 4, we will layout theorems for the ordinary least squares (ols) and Lasso case to show that the procedure adapts to different situations with different asymptotic properties. Second, the estimator can be interpreted as a \textquotedblleft soft clustering\textquotedblright algorithm, as each individual has not to be assigned to one group ($0$-$1$-weights), but allows flexible weights and continuum number of groups. By this soft-clustering of individuals whose functions or coefficents are close enough, each second-stage estimator will be more accurate in terms of the mean squared error (MSE). This is  because the variance of the estimation reduces as compared to the individual estimator.
\end{remark}


In the following we will specify the generic algorithm for linear panel data models and estimate them with ols in the low-dimensional setting and with lasso in the high-dimensional setting.



\subsection{Linear Panel Data Model}
\label{sec:ADS_low}

To begin with, let us first consider the linear panel data model from Example 2 \eqref{eq:Lasso} in a low-dimensional setting, i.e.  where the dimensionality $p$ of each predictor $x_{it}$ is fixed and independent of the sample size. We illustrate our ADS estimator for this case based on the OLS estimation.


\begin{algorithm}[ADS-Algorithm for linear panel data]
~

\begin{itemize}
\item[(1)](First stage estimation) Using the OLS estimator to construct the initial individual estimators:

    \[
        \hat \beta^0_i = \argmin_{\beta\in \mathbb{R}^p} \sum_{t=1}^T(y_{it}-x_{it}' \beta)^2 \quad \text{for} \quad i=1,\ldots, N.
    \]
    \item[(2)](Constructing the weight matrix) Compute the $N \times $N weight matrix $W$  as follows,
    \begin{equation}\label{eq:Wseries}
    W(i,j):=\delta \exp(-\gamma \|\hat \beta^0_i- \hat \beta^0_j\|_2^2),
    \end{equation}
    for all $i$ and $j$.
    \item[(3)](Second stage estimation) Re-estimate each $\beta_i^*$ by the weighted OLS:
     \[
     \tilde \beta_i:=\argmin_{\beta\in \mathbb{R}^{p}} \sum_{j=1}^N\sum_{t=1}^T{W(i,j)(y_{jt}-x_{jt}'\beta)^2}  \quad \text{for} \quad i=1,\ldots, N.
\]
\end{itemize}
\label{Algo:linear}
\end{algorithm}

When the sample size $N$ and $T$ are large, the choice of $\delta$ in \eqref{eq:Wseries} does not matter for the asymptotic performance of the ADS procedure. In practice, we could simply choose a $\delta\in (0,1)$. For example, in our study,  
the $\delta$ is set to be $0.5$. We further make a few important comments about the algorithm.


\begin{remark}[Extension to an iterative approach]
As mentioned earlier, the algorithm can be extended by iterating the steps to update first the initial estimates and then the weighting matrix. More precisely for the ols estimtaor, after obtaining $\widetilde{\beta}_i$ from Step 3, we could update the weight matrix in Step 2 using $\widetilde{\beta}_i$ and repeat Step 3 to obtain refined estimators. One could do either a one-step correction or repeat Step 2 and Step 3 for several iterations until the change of the weights falls below a certain threshold.
\end{remark}

\subsection{High-dimensional Panel Data Models}

One interesting special case of the Algorithm \ref{Algo:generic} is the high-dimensional linear regression case which is the core contribution of the paper. As high-dimensional data are more and more available for researchers, we would like to work out a special version of Algorithm \ref{Algo:generic} for Lasso estimation of panel data in this subsection.

For high-dimensional linear regression, many estimators have been proposed in literature, such as Lasso \citep{Tibshirani:96}, Dantzig selector \citep{CandesETao2007}, square-root Lasso \citep{BCW-SqLasso} (or scaled Lasso \citep{SunZhang:12}),  $L_2$ Boosting \citep{LS:2016}, and post-Lasso \citep{BCPostLASSO}. In principle, our ADS algorithm can be combined with any estimator for high-dimensional linear regression. For the ease of illustration, we will adopt the most widely used Lasso estimator and present the high-dimensional ADS algorithm next.


Consider the following high-dimensional model: for $1\leq i \leq N$ and $1 \leq t \leq T$,
\begin{equation}
y_{it}=x_{it}^T \beta_i^* +\varepsilon_{it},
\end{equation}
where each $\varepsilon_{it}$ are i.i.d. errors draw from a normal distribution with mean zero and variance $\sigma^2$, $N(0, \sigma^2)$.

\begin{remark}
The distributional assumption on the error terms can be relaxed. For example, serial correlation can be introduced or heteroskedasticity by employing the theory of self-normalized processes for Lasso as proposed in \cite{BCCH12}. 
\end{remark}

For each $\beta_i^* \in \mathbb{R}^p$ we allow $p \gg T$, and $p$ has to be of order $O(\exp(T^\alpha))$ for some $\alpha\in (0,1)$. Each $\beta_i^*$ depends on the unobserved variable $\alpha_i$, i.e. $\beta_i^*=\beta(\alpha_i)$ where $\beta: \mathbb{R} \rightarrow \mathbb{R}^p$ and $\alpha_i$ is a scalar. This can be interpreted in different ways as pointed out earlier.

\begin{algorithm}[ADS-Algorithm for Lasso]\label{Algo:Lasso}

\begin{itemize}
	\item[(1)] Run $N$ separate Lassos:
	\[
	 \hat \beta_{i, \text{Lasso}}:=\\argmin_{\beta_i} \frac{1}{2T}\sum_{t=1}^T  (y_{it}-x_{it}^T \beta_{i})^2 +\lambda_i \|\beta_i\|_1,
	\]
    where $\lambda_i$ is the penalty loading for individual $i$ in the first stage.
	\item[(2)]Compute $W$ matrix such that
    \begin{equation}\label{eq:W}
    W(i,j):=\delta exp(-\gamma || \hat \beta_{i, \text{Lasso}}- \hat \beta_{j, \text{Lasso}}||_2^2),
    \end{equation}
    for all $i\neq j$. $W(i,i)=1$ for all $i=1,2,...,N$.
	\item[(3)] Estimation the final estimator $\widetilde{\beta}_i$ for $1 \leq i \leq N$:
 \[ \widetilde{\beta}_i:=\\argmin_{\beta_i \in \mathbb{R}^p} \frac{1}{2T\widetilde{N}_i} \sum_{j=1}^N W(i,j) \sum_{t=1}^T (y_{jt}-x_{jt}^T\beta_i)^2 + {\lambda}_{2,i}\|\beta_i\|_1,
\]
where $\widetilde{N}_i=\sum_{j=1}^N W(i,j)$, and $\widetilde{\lambda}_i$ is the penalty loading for individual $i$ in the second stage.
\end{itemize}
\end{algorithm}


For the high dimensional data, in Algorithm \ref{Algo:Lasso}, we can replace  Lasso with other statistical or machine learning procedures that work well in high-dimensions, e.g., $L_2$ Boosting (\cite{LS:2016}), Post-Lasso (\cite{BCPostLASSO}). For brevity, we shall not repeat these procedures in this paper and restrict to the Lasso case.

\section{Simulation Study}

In this section we present results from a simulation study, in particular we compare our methods to the \textquotedblleft frequentist\textquotedblright approach, i.e. estimating individual functions with only observations from this individual. We present results for OLS and Lasso regression.


\subsection{Linear Regression}

We consider the following model

\begin{equation}
y_{it} = \beta_i' x_{it} + \varepsilon_{it}, i=1,\ldots,n, t=1,\ldots,T,
\label{LR}
\end{equation}

with $\varepsilon_{it} \sim N(0, \sigma^2)$ iid and $x_i=(x_{i1}, \ldots, x_{iT})'$ ($T \times (p+1)$ matrix) with $x_{it}$ a $(p+1)-$dimensional vector containing a constant of $1$. For $(x_{it2}, \ldots, x_{it(p+1)})$ we consider two settings:
\begin{enumerate}
	\item iid setting: $(x_{it2}, \ldots, x_{it(p+1)}) \sim N(0, \sigma^2 I_{p})$.
	\item correlated setting: $(x_{it2}, \ldots, x_{it(p+1)}) \sim N(0, \Sigma)$ where $\Sigma$ has Toeplitz structure with parameter $0.5$.
\end{enumerate}

We consider two data generating processes, one specifies the $(p+1)-$dimensional coefficient vector $\beta_i$ as a function of an random, unobserved quantity $\alpha_i$ and one assumes that the $\beta_i$ are drawn from a normal distribution with constant correlation $\rho$ between the same component of different observations $i$ but independent components.

\subsubsection{DGP 1}\label{setting-1}
The individual coefficients
$\beta_i$ ($p+1$-dimensional) are simulated
in such a way -- based on the normal distribution -- that the components are
uncorrelated for individual $i$, but the (same) components are
correlated between individuals $i$ and $j$ with correlation
$\rho$. Namely, $cor(\beta_{ik}, \beta_{il})=0$ for $k \neq l$ and
$cor(\beta_{ik}, \beta_{jk})=\rho$ for $i\neq j$.
So, $(\beta_{1l}, \ldots, \beta_{Nl}) \sim \mathcal{N}(0,\Sigma)$. The $N-$dimensional correlation matrix has entries $\rho$ off the diagonal. The vectors are iid for different components $l$ $(l=1,\ldots,p+1)$.

\subsubsection{DGP 2}\label{setting-2}

This DGP is similar to 1,  but $\beta_i$ is generated differently, namely:
\[
\beta_i'=(1,\ldots,1)' + (\alpha_i, \alpha_i^2, -\alpha_i, - \alpha_i^2,\alpha_i, \alpha_i/2, \alpha_i/3, \ldots).
\]
The $\alpha_i$ are drawn iid from a uniform distribution, $\alpha_i \sim \mathcal{U}(0,1)$.

\subsection{Lasso Estimation / Machine Learning}

Here we estimate a model comparable to \ref{LR}, but we use Lasso and Lasso with discrete smoothing for estimation. Lasso usually assumes a sparse setting, i.e. it is assumed that there is a set of $p+1$ potential variables, but only a small subset of $s+1$ regressors has non-zero coefficients. As data generating processes we use variants of DGP 1 and DGP 2 which are described above: the first $s+1$ components are simulated in exactly the same way as in DGP1 and DGP2 and the remaining $p-s$ variables are set equal to zero. In the simulations we vary $n,p,s$ and $T$.
We decompose the vector $\beta_i=(\beta_i^1, \beta_i^2)$ in two parts: $\beta_i^1=(\beta_{i1}, \ldots, \beta_{i(s+1)})$ and $\beta_i^2=(0,\ldots,0).$ Hence we can formulate the DGPs in the following way
\begin{itemize}
	\item[DGP 3] $\beta_i^1 = (1,\ldots,1)' + (\alpha_i, \alpha_i^2, -\alpha_i, -\alpha_i^2, \alpha_i/1, \alpha_i/2, \alpha_i/3, \ldots)'$ and $\alpha_i \sim \mathcal{U}[0,1]$ iid 
	\item[DGP 4] For each component $j$: $\beta^1_{ij}, i=1,\ldots,n$ are multivariate normal distributed with constant correlation $\rho$, $N(0,\Sigma)$ with $\Sigma$ a $n\times n$ correlation matrix with unit correlation on the diagonal and off-diagonal entries of $\rho$. The components are independent.
\end{itemize}

\subsection{Results}\label{results}
In this section we present the simulation results. We generate the data according to the DGPs described above and estimate it with OLS (DGP 1 and 2) or Lasso  (DGP 3 and 4) and the smoothing version introduced in this paper.
We vary different parameters. For estimation we use data set of size $n,T$. The forecasts are evaluated out-of-sample (same sample size as used for estimation) according to the mean squared error (MSE). We set the number of repetitions to $R=500$. We use a setting where the design matrix $X$ is iid (``X iid'') and setting where $X$ is correlated (``X corr''), i.e.~has Toeplitz structure. In the linear case we set the number of parameters $N,T,p$. In the Lasso
case we use a exact sparsity design, where $p$ denotes the number of parameters and $s$ the number of non-zero coefficients.

The results can be summarized as follows: when $T$ is small and $N$ is large our methods performs very favorable. When $N$ is small and $T$ is large, discrete smoothing does not add much benefit. This is exactly what one would expect. Overall, the simulation results highlight the usefulness of our approach for empirical applications.

\newpage

\subsubsection{OLS Setting}
\textcolor{white}{.} 
\begin{longtable}{@{}lllllll@{}}
	\caption{Simulation Results Linear - out of sample - Setting 1, X iid}\tabularnewline
	\toprule
	  &   &     & \multicolumn{2}{l}{p=5} & \multicolumn {2}{l}{p=10}  \tabularnewline
	N & T & cor & MSE OLS & MSE DS          & MSE OLS & MSE DS  \tabularnewline
	\midrule
	\endfirsthead
	\toprule
	  &   &     & \multicolumn{2}{l}{p=5} & \multicolumn {2}{l}{p=10}  \tabularnewline
	N & T & cor & MSE OLS & MSE DS          & MSE OLS & MSE DS  \tabularnewline
	\midrule
	\endhead
	2 & 10 & 0 & 2.0477 & 1.8037 & & \tabularnewline
	2 & 10 & 0.3 & 1.8083 & 1.5804& &\tabularnewline
	2 & 10 & 0.7 & 1.8056 & 1.3903& &\tabularnewline
	2 & 10 & 1 & 2.0681 & 1.1047& &\tabularnewline
	10 & 10 & 0 & 1.8735 & 1.3439& &\tabularnewline
	10 & 10 & 0.3 & 1.8975 & 1.2038& &\tabularnewline
	10 & 10 & 0.7 & 1.8734 & 0.8594& &\tabularnewline
	10 & 10 & 1 & 1.9841 & 0.2563& &\tabularnewline
	50 & 10 & 0 & 1.9364 & 1.6732& &\tabularnewline
	50 & 10 & 0.3 & 1.8925 & 1.5160& &\tabularnewline
	50 & 10 & 0.7 & 1.9256 & 0.9738& &\tabularnewline
	50 & 10 & 1 & 1.8750 & 0.0557& &\tabularnewline
	2 & 20 & 0 & 0.4585 & 0.4601 & 1.4150 & 1.3666\tabularnewline
	2 & 20 & 0.3 & 0.4429 & 0.4380 & 1.3444 & 1.2612\tabularnewline
	2 & 20 & 0.7 & 0.4405 & 0.4027 & 1.3695 & 1.1300\tabularnewline
	2 & 20 & 1 & 0.4509 & 0.2285 & 1.3662 & 0.6643 \tabularnewline
	10 & 20 & 0 & 0.4501 & 0.6132& 1.3668 & 1.1860 \tabularnewline
	10 & 20 & 0.3 & 0.4505 & 0.6299 & 1.3642 & 1.1461\tabularnewline
	10 & 20 & 0.7 & 0.4605 & 0.5729& 1.3681 & 0.9733\tabularnewline
	10 & 20 & 1 & 0.4546 & 0.0444 &1.3442 & 0.1208\tabularnewline
    50 & 20 & 0 && & 1.3651 & 1.6518\tabularnewline
    50 & 20 & 0.3 && & 1.3691 & 1.7681\tabularnewline
    50 & 20 & 0.7 && & 1.3730 & 1.4962\tabularnewline
    50 & 20 & 1 && & 1.3598 & 0.0211\tabularnewline
    2 & 50 & 0 && & 0.2896 & 0.2901\tabularnewline
    2 & 50 & 0.3 && & 0.2890 & 0.2890\tabularnewline
    2 & 50 & 0.7 && & 0.2942 & 0.3010\tabularnewline
    2 & 50 & 1 && & 0.2805 & 0.1429\tabularnewline
    10 & 50 & 0 && & 0.2907 & 0.3431\tabularnewline
    10 & 50 & 0.3 && & 0.2888 & 0.4161\tabularnewline
    10 & 50 & 0.7 && & 0.2856 & 0.5958\tabularnewline
    10 & 50 & 1 && & 0.2868 & 0.0270\tabularnewline
	\bottomrule
\end{longtable}

\begin{table}[!htb]
	\caption{Simulation Results Linear - out of sample - Setting 1, X cor,	p=5}\tabularnewline
\begin{minipage}{0.11\textwidth}
	\addtocounter{table}{-1}
\begin{longtable}[l]{@{}lllll@{}}
	\toprule
	N & T & cor & MSE OLS & MSE DS\tabularnewline
	\midrule
	\endfirsthead
	\toprule
	N & T & cor & MSE OLS & MSE DS\tabularnewline
	\midrule
	\endhead
	2 & 10 & 0 & 1.8411 & 1.6828\tabularnewline
	2 & 10 & 0.3 & 1.9002 & 1.6448\tabularnewline
	2 & 10 & 0.5 & 1.9717 & 1.6382\tabularnewline
	2 & 10 & 0.8 & 1.9106 & 1.4015\tabularnewline
	2 & 10 & 1 & 1.9148 & 1.1648\tabularnewline
	10 & 10 & 0 & 1.9718 & 1.4550\tabularnewline
	10 & 10 & 0.3 & 1.9239 & 1.2699\tabularnewline
	10 & 10 & 0.5 & 1.9413 & 1.1295\tabularnewline
	10 & 10 & 0.8 & 2.1294 & 0.9679\tabularnewline
	10 & 10 & 1 & 1.9357 & 0.3262\tabularnewline
	50 & 10 & 0 & 1.9365 & 1.6522\tabularnewline
	50 & 10 & 0.3 & 1.9011 & 1.4880\tabularnewline
	50 & 10 & 0.5 & 1.9335 & 1.2989\tabularnewline
	\bottomrule
\end{longtable}
\end{minipage}\qquad 
\begin{minipage}{0.7\textwidth}
	\addtocounter{table}{-1}
	\begin{longtable}[r]{@{}lllll@{}}
		\toprule
		N & T & cor & MSE OLS & MSE DS\tabularnewline
		\midrule
		\endfirsthead
		\toprule
		N & T & cor & MSE OLS & MSE DS\tabularnewline
		\midrule
		\endhead
	
		50 & 10 & 0.8 & 1.8970 & 0.7490\tabularnewline
		50 & 10 & 1 & 1.8919 & 0.0782\tabularnewline
		100 & 10 & 0 & 1.9235 & 1.9808\tabularnewline
		100 & 10 & 0.3 & 1.9332 & 1.7589\tabularnewline
		100 & 10 & 0.5 & 1.9416 & 1.4893\tabularnewline
		100 & 10 & 0.8 & 1.9211 & 0.8225\tabularnewline
		100 & 10 & 1 & 1.8920 & 0.0485\tabularnewline
		2 & 20 & 0 & 0.4459 & 0.4596\tabularnewline
		2 & 20 & 0.3 & 0.4490 & 0.4546\tabularnewline
		2 & 20 & 0.5 & 0.4526 & 0.4351\tabularnewline
		2 & 20 & 0.8 & 0.4789 & 0.4131\tabularnewline
		2 & 20 & 1 & 0.4717 & 0.2574\tabularnewline
		\\
		\bottomrule
	\end{longtable}
\end{minipage}
\end{table}

\begin{table}[t]
\begin{longtable}[]{@{}llllll@{}}
	\caption{Simulation Results Linear - out of sample - Setting 2, p=5}\tabularnewline
	\toprule
	   &   & \multicolumn{2}{l}{X iid} & \multicolumn {2}{l}{X cor}  \tabularnewline
	 N & T & MSE OLS & MSE DS          & MSE OLS & MSE DS \tabularnewline
	\midrule
	\endfirsthead
	\toprule
  	  &       & \multicolumn{2}{l}{X idd} & \multicolumn {2}{l}{X cor}  \tabularnewline
	N & T     & MSE OLS & MSE DS          & MSE OLS & MSE DS  \tabularnewline
	\midrule
	\endhead
	2 & 10 & 1.7442 & 1.1173& 1.8793 & 1.2295\tabularnewline
	5 & 10 & 1.9357 & 0.6869& 1.9473 & 0.7309\tabularnewline
	10 & 10 & 1.9140 & 0.4323 & 1.9075 & 0.4363\tabularnewline
	50 & 10 & 1.9496 & 0.2680& 1.9488 & 0.2444\tabularnewline
	2 & 20 & 0.4488 & 0.3057 & 0.4574 & 0.3079\tabularnewline
	5 & 20 & 0.4558 & 0.1996 &  0.4475 & 0.2006\tabularnewline
	10 & 20 & 0.4630 & 0.1653 & 0.4481 & 0.1643\tabularnewline
	50 & 20 & 0.4520 & 0.1561& 0.4557 & 0.1546\tabularnewline
	2 & 50 & 0.1416 & 0.1155& 0.1395 & 0.1140\tabularnewline
	5 & 50 & 0.1402 & 0.1071 & 0.1395 & 0.1064\tabularnewline
	10 & 50 & 0.1411 & 0.1160& 0.1378 & 0.1136\tabularnewline
	50 & 50 & 0.1375 & 0.1298& 0.1389 & 0.1310\tabularnewline
	2 & 100 & 0.0640 & 0.0663& 0.0636* & 0.0655*\tabularnewline
	\bottomrule
\end{longtable}
\end{table}
\subsubsection{Lasso Setting}
\textcolor{white}{.} 
\begin{longtable}[!b]{@{}lllllll@{}}
	\caption{Simulation Results Lasso - out of sample - Setting 1, 
		p=15, s=5}\tabularnewline
	\toprule
	  &   &     & \multicolumn{2}{l}{X iid} & \multicolumn {2}{l}{X cor}  \tabularnewline
	N & T & cor & MSE OLS & MSE DS          & MSE OLS & MSE DS \tabularnewline
	\midrule
	\endfirsthead
	\toprule
	  &   &     & \multicolumn{2}{l}{X iid} & \multicolumn {2}{l}{X cor}  \tabularnewline
	N & T & cor & MSE OLS & MSE DS          & MSE OLS & MSE DS \tabularnewline
	\midrule
	\endhead
	10 & 10 & 0 & 6.5757 & 5.5606& 3.5783 & 3.1561\tabularnewline
	10 & 10 & 0.5 & 4.8400 & 2.5980& 3.7155 & 2.3549\tabularnewline
	10 & 10 & 1 & 5.4107 & 1.6479 & 5.0942 & 1.8677\tabularnewline
	50 & 10 & 0 & 10.3866 & 9.6920& 11.2201 & 11.0931\tabularnewline
	50 & 10 & 0.5 & 10.9481 & 8.6209 & 4.4456 & 2.9577\tabularnewline
	50 & 10 & 1 & 5.6059 & 1.2032& 3.9552 & 0.2987\tabularnewline
	10 & 25 & 0 & 1.4707 & 1.5144& 1.3949 & 1.5799\tabularnewline
	10 & 25 & 0.5 & 1.3978 & 1.0465& 1.3634 & 1.1555\tabularnewline
	10 & 25 & 1 & 1.4948 & 0.0687& 1.4258 & 0.1219\tabularnewline
	50 & 25 & 0 & 1.4107 & 2.3890& 1.3604 & 2.4301\tabularnewline
	50 & 25 & 0.5 & 1.4153 & 1.5704& 1.3976 & 1.6055\tabularnewline
	50 & 25 & 1 & 1.4239 & 0.0114 & 1.3339 & 0.0150\tabularnewline
	\midrule
	100 & 25 & 0 & 1.4205 & 2.8330& 1.3944 & 2.9070\tabularnewline
	100 & 25 & 0.5 & 1.3933 & 1.8057& 1.3499 & 1.7905\tabularnewline
	100 & 25 & 1 & 1.3840 & 0.0055& 1.3184 & 0.0079\tabularnewline
	10 & 50 & 0 & 0.5047 & 0.6573& 0.6541 & 0.8595\tabularnewline
	10 & 50 & 0.5 & 0.4862 & 0.6352& 0.6403 & 0.7694\tabularnewline
	10 & 50 & 1 & 0.4678 & 0.0232& 0.6099 & 0.0407\tabularnewline
	50 & 50 & 0 & 0.4920 & 1.6252& 0.6430 & 1.8651\tabularnewline
	50 & 50 & 0.5 & 0.4924 & 1.3506& 0.6494 & 1.4133\tabularnewline
	50 & 50 & 1 & 0.4595 & 0.0039& 0.5710 & 0.0064\tabularnewline
	100 & 50 & 0 & 0.4951 & 2.1594& 0.6376 & 2.3625\tabularnewline
	100 & 50 & 0.5 & 0.4879 & 1.6079  & 0.6496 & 1.6408\tabularnewline
	100 & 50 & 1 & 0.5212 & 0.0021& 0.6619 & 0.0030\tabularnewline
	\bottomrule
\end{longtable}

\begin{longtable}[]{@{}llllll@{}}
	\caption{Simulation Results Lasso - out of sample - Setting 2, p=15, s=5}\tabularnewline
	\toprule
	  &   & \multicolumn{2}{l}{X iid} & \multicolumn {2}{l}{X cor}  \tabularnewline
	N & T & MSE OLS & MSE DS          & MSE OLS & MSE DS \tabularnewline
	\midrule
	\endfirsthead
	\toprule
	  &   & \multicolumn{2}{l}{X iid} & \multicolumn {2}{l}{X cor}  \tabularnewline
	N & T & MSE OLS & MSE DS          & MSE OLS & MSE DS \tabularnewline
	\midrule
	\endhead
	 2 & 10 & 6.5041 & 4.9281 & 5.7229 & 4.1633\tabularnewline
	 5 & 10 & 6.8800 & 2.6220& 5.7531 & 1.7605\tabularnewline
	 10 & 10 & 6.8314 & 1.0121 & 5.5935 & 0.7353\tabularnewline
	 50 & 10  & 6.8359 & 0.3740 & 5.5829 & 0.2970\tabularnewline
	 2 & 20 & 2.9264 & 1.5121  & 1.6070 & 0.7536\tabularnewline
	 5 & 20 & 3.0506 & 0.4947& 1.5618 & 0.2914\tabularnewline
	 10 & 20 & 2.9579 & 0.2814& 1.5467 & 0.2080\tabularnewline
	 50 & 20 & 2.9423 & 0.2257 & 1.5342 & 0.1651\tabularnewline
	 2 & 50 & 0.3853 & 0.1905 & 0.1816 & 0.1330\tabularnewline
	 5 & 50 & 0.3577 & 0.1351& 0.1751 & 0.1111\tabularnewline
	 10 & 50 & 0.3592 & 0.1422 & 0.1771 & 0.1174\tabularnewline
	 50 & 50 & 0.3782 & 0.1684  & 0.1777 & 0.1288\tabularnewline
	\bottomrule
\end{longtable}

\newpage

\section{Empirical Studies: DIDI data and gap prediction}
\subsection{Data Set}
Didi is a major ride-sharing company in China with more than $400$ million users across $400$ cities. As the services are based on smartphone applications the company can collect a huge amount of data on a daily basis concerning detailed information on asked and provided rides including detailed geographical information and time stamps. We use the raw data on single rides to estimate the gap in demand which is defined as the difference of requested rides, i.e. the number of calls in a certain location, within a period of 10 minutes and the number of answers in this location, within this period. The gap is an important variable for the company as good predictions of this variable help to hold available drivers and it serves as input for dynamic pricing strategies.
The data set contains the following variables: 
\begin{itemize}
	\item location: the location id  corresponds to a clustered map, dividing cities in hexagons (66).
	\item time (Date\_str, DayM, DayW, TimeP): date, day of month, day of week (1-7), time period of the day (1-144).
  \item weather: categorical variable on weather conditions
  \item temperature in a certain location at a certain point of time
	\item measure of air pollution (continuous)
	\item requests (demand): number of calls in a certain location in time spans of 10 minutes
	\item answers (supply): number of answers in a certain location in time spans of 10 minutes
	\item gap defined as the difference of requests and answers
	\item road / traffic conditions (four variables, L1-L4)
	\item price: the average price/ median price within a 10-minute-cell at a certain location, computed from the raw data.
\end{itemize}


We build a model to predict the gaps on location level, i.e. for each district we estimate a model for gap. It seems plausible that the demand depends on the location (e.g. city center vs. outskirt). This leads to $66$ individual location units in the city we consider for our analysis. The total sample size of observations for all units is $n=135,674$. We split the data set in a training sample ($n=105,000$) and a testing sample ($n=30,674$) under consideration of the location structure. So to estimate each model appr. 1,500 observations are available.\footnote{For each unit the last days of the period of record are removed.} We model variables like day of month and day of week as categorical leading to a large set of potential covariates for each unit, namely $46$ which is challenging. This situation which is common for many real word applications, in particular data sets collected on the internet fits well for the methods we proposed as additional information from other observations is very valuable in this setting.

\subsection{Results}

We estimate for each unit linear regression models with OLS and Lasso and compare the results with the adaptive discrete smoothing method for each estimator introduced in this paper. Although the OLS estimator is not well defined for units with $p>n$, it can be used for prediction.

We apply the ADS procedure to didichuxing.com's data. This data involves 66 districts. In our study, we found that our ADS procedure outperforms all other methods, measured by both in-sample and out-of-sample MSE. Already in the simulation study we found clear evidence that the ADS procedure does better in MSE compared to traditional methods, in many different settings.

\begin{table}
\begin{tabular}{ll}
	\hline
   Method & MSE\\ \hline
   naive estimator &$1.37$ \\
   OLS & $2.25$ \\ 
   Lasso & $1.19$ \\
	 ADS OLS &$1.18$\\
	ADS Lasso &$1.00$\\
	\hline
 \end{tabular}
\end{table}

\section{Conclusion}

In this paper, we propose a novel adaptive discrete smoothing procedure (ADS). Such a procedure is applicable to both non-linear and high-dimensional panel data.\footnote{The core idea of the procedure can also applied to many other statistical problems. An application to non-parametric estimation with discrete and continuuos regressors is given in \cite{CLS2020}.} The procedure is especially useful in the large cross-sectional size and small longitude scenario, where building individual models based on individual data is highly imprecise, while heterogeneity across cross-sectional individuals prevents good performance of a uniform modeling process for all individuals. The results of our simulation strongly support our procedure compared to other existing econometrics or statistics methodologies. Our procedure can also be interpreted as a clustering method. In future research we would like to derive the theoretical properties (work in progress) and explore more formally the connection to cluster methods.

\pagebreak
\bibliographystyle{aea}
\bibliography{mybib}

\end{document}